\documentclass{article}

\usepackage{microtype}
\usepackage{graphicx}
\usepackage{subfigure}
\usepackage{booktabs} 

\usepackage{hyperref}

\usepackage[accepted]{icml2020}

\begin{document}

\twocolumn[
\icmltitle{Bringing the People Back In: Contesting Benchmark Machine Learning Datasets}



\icmlsetsymbol{equal}{*}

\begin{icmlauthorlist}
\icmlauthor{Emily Denton}{equal,google}
\icmlauthor{Alex Hanna}{equal,google}
\icmlauthor{Razvan Amironesei}{usf}
\icmlauthor{Andrew Smart}{google}
\icmlauthor{Hilary Nicole}{google}
\icmlauthor{Morgan Klaus Scheuerman}{google}
\end{icmlauthorlist}

\icmlaffiliation{google}{Google Research}
\icmlaffiliation{usf}{University of San Francisco}

\icmlcorrespondingauthor{Emily Denton}{dentone@google.com}
\icmlcorrespondingauthor{Alex Hanna}{alexhanna@google.com}

\icmlkeywords{Machine Learning, ICML}

\vskip 0.3in
]



\printAffiliationsAndNotice{\icmlEqualContribution} 

\begin{abstract}
In response to algorithmic unfairness embedded in sociotechnical systems, significant attention has been focused on the contents of machine learning datasets which have revealed biases towards white, cisgender, male, and Western data subjects. In contrast, comparatively less attention has been paid to the histories, values, and norms embedded in such datasets. In this work, we outline a research program -- a {\it genealogy} of machine learning data -- for investigating how and why these datasets have been created, what and whose values influence the choices of data to collect, the contextual and contingent conditions of their creation. We describe the ways in which benchmark datasets in machine learning operate as infrastructure and pose four research questions for these datasets. This interrogation forces us to ``bring the people back in'' by aiding us in understanding the labor embedded in dataset construction, and thereby presenting new avenues of contestation for other researchers encountering the data.  
\end{abstract}

\section{Introduction}
\label{intro}
Sociotechnical systems abound in ways that they have failed people of color \citep{noble2018algorithms, benjamin2019race}, women \citep{Bolukbasi2016}, LGBTQ+ communities \citep{Scheuerman2019}, people with disabilities \citep{Hutchinson2020, Trewin2018AIFF}, and the working class and those in poverty \citep{Eubanks2018}.  Many of these failures have been attributed to under-representation of these groups in the data upon which these systems are built or undesirable correlations between certain groups and target labels in a dataset. In response, a proliferation of algorithmic fairness interventions have emerged that hinge on parity of representation of different demographic groups within training datasets. While interventions of this sort play a non-trivial role in achieving recently advanced technical definitions of algorithmic fairness (e.g. \citet{Hardt2016}), failures of data-driven systems are not located exclusively at the level of those who are represented or under-represented in the dataset. Furthermore, data collection efforts aimed at increasing the representation of marginalized groups within training data are too often carried out through exploitative or extractive mechanisms mechanisms \citep{DIFnbc}.


In contrast to the significant efforts focused on statistical properties of training datasets, comparatively little attention has been paid to how and why these datasets have been created, what and whose values influence the choices of data to collect, the contextual and contingent conditions of their creation, and the emergence of current norms and standards of data practice. 

In this work, we motivate and proposed a research program for constructing a {\it genealogy of data} applied to benchmark machine learning datasets. Our research program adopts Michel Foucault's method of genealogy \citep{foucault1977}, an interpretive method that traces the historical formation and transformation of practices, discourses, and concepts. Our work is motivated, in large part, by \citeauthor{crawford2019excavating}'s {\it archaeology} of several computer vision datasets, an endeavor aimed at exposing the assumptions and values underlying prominent machine learning datasets \citeyearpar{crawford2019excavating}. Our work is similarly concerned with ethical and political dimensions of what has been taken-for-granted in dataset construction, the ontologies that structure prominent datasets, and the epistemic commitments that are often (invisibly) embedded in datasets and data practices. Through studying data artifacts and surrounding discourses, our genealogy further aims to trace the emergence of the shared work practices that structure the development and use of machine learning datasets.

This research program centers on ``bringing the people back in'' to the study of datasets used in the training of machine learning systems. Bringing the people back in forces us to focus on the contingent, historical, and value-laden work practices of actual machine learning researchers.  Moreover, opening this box is not merely an avenue towards more transparency, although this is a necessary first step.  As \citet{mulligan2019shaping} note, focusing on transparency with the goal of showing the internals of a system without plausible actions of being able to change aspects of that system are a Pyrrhic victory. Contestability, however, allows us to critically engage within the system and provides us with the ability to "iteratively identify and embed domain knowledge and contextual values" into such a system. We aim to help flesh out the unspoken labor which goes into the creation of datasets to provide new avenues into contestability of these important information infrastructures. 


Our primary contributions in this work as are follows. First, we introduce a new vocabulary and concepts from infrastructural studies to frame out understanding of data with respect to modes of power and contestability. In doing so, we motivate the need for genealogical method to trace the histories of, and de-naturalize, this data infrastructure. We then outline the components of a novel research program for a genealogy of machine learning data and end by summarizing our forward-looking goals.

\section{Data Infrastructure}
In this work, we situate our understanding of data within the conceptual framework of infrastructure, arguing that datasets -- as well as the practices surrounding the development and use of such datasets -- operate as a form of infrastructure for machine learning research and development. 

We use infrastructure in a broad sense, to encompass the conceptual and material tools that enable different forms of knowledge work and scientific practice, echoing the definition from infrastructure studies \citep{sorting, Bowker2010, Larkin2013ThePA}. Infrastructure is characterized, we argue, by a set of core features: it is embedded into, and acts as the foundation, for other tools and technologies; when working as intended for a particular community, it tends to seep into the background and become incorporated into routines; the invisibility of infrastructure, however, is situated - what is natural or taken for granted from one perspective may be highly visible or jarring from another; though frequently naturalized, infrastructure is built, and thus inherently contextual, situated, and shaped by specific aims. 

So, in what sense do datasets operate as infrastructure? 
At the most obvious and localized level, training datasets determine what a resulting machine learning model learns, how problems are framed, and what solutions are prioritized. Statistical properties of a dataset
determine category boundaries and who/what is rendered legible by a downstream model. Furthermore, labelled datasets organized by a particular categorical schema frequently subsume modeling decisions regarding the conceptualization, operationalization, and measurement of target variables for downstream classification systems and datasets frequently embed metrics of success.

Second, datasets play a significant role in benchmarking AI algorithms. Benchmark datasets that are recognized as go-to standards for evaluation and comparison often take on an authoritative role and improvements on performance metrics associated with the benchmark become synonymous with progress in the subfield. Datasets that have achieved such authoritative status also play a unique and powerful role in structuring research agendas and values within machine learning subfields \citep{Dotan2020ValueladenDS}.

Third, because datasets and their associated benchmarks take on this authoritative nature within machine learning, they often take the status of the ``model organism'' within laboratory studies. The characteristics of the model organism are pragmatic: readily available, easy to manipulate, and somewhat uncomplicated in form. However, the cheapness and availability of the model organism also open itself up to a set of conceptual and empirical gaps. For instance, in her critique of Twitter as one of the most common model organisms, the fruit fly (or {\it drosophila melanogaster}) of large-scale social media research, \citet{tufekci2014big} points to how such a focus obscures more complicated social processes at work, as the particular technological affordances of the platform and its niche user population become a stand-in for those processes. Datasets and authoritative benchmarks, then, with their contingent collection processes, annotation and archival practices become a stand-in for more complicated data traces and machine learning tasks.

Fourthly and finally, publicly available research datasets act as infrastructure by providing the methodological backbone of how AI tools are deployed in industry contexts. The boundary between research and practice is thin and pliable, as AI researchers flit between academia and industry. Accordingly, that research follows them and enters into commercial products. Most technology companies derive value from the amount and kind data they collect, and those data are much larger than those publicly available research datasets. However, these shifts are conceptualized by researchers as merely changes in scale and rarely in kind. These datasets perform an infrastructural function by undergirding the material research needs upon which commercial AI is also built and deployed.

Working infrastructure tends to become invisible and naturalized within everyday routines. The concept of naturalization provides language with which to describe the dominant data practices within the field of machine learning. For example, countless subjective and value-laden decisions go into the construction of a dataset.  Yet, once a dataset is released and becomes established enough to seamlessly support research and development, the contingent conditions of creation tend to be lost or taken for granted. Once naturalized, datasets are more likely to be treated as neutral or scientific objects and uncritically adopted within daily work routines. 

The norms and standards that structure data is collection and use have also become naturalized to an extent that they are frequently taken for granted by machine learning practitioners. This is exemplified by the limited focus on -- and often complete absence of --  data considerations within machine learning textbooks and curriculum (e.g. \citet{Goodfellow-et-al-2016}), the under-specification or data decisions in publications accompanying new datasets \cite{Geiger2020, Scheuerman2020}, and the routine undervaluing of the work that goes into the construction of datasets \cite{nlp_clever_hans, archives}.

Though frequently naturalized or taken for granted, infrastructure is built, and thus inherently contextual, situated, and shaped by specific aims.  By attending to the way in which data infrastructure is built and maintained our genealogy provides an avenue of "bring the people back in" to the analysis of datasets. We are also reminded that the very notion of ‘working infrastructure’ is contingent on perspective -- the background operating conditions for one person may be a daily object of concern for another \citep{Larkin2013ThePA}.




By tracing the histories and contingent conditions of creation of datasets and data practices, we seek to make visible and thus de-naturalize data infrastructure. In this sense, our genealogy of data follows the the methodological theme of infrastructural inversion \cite{Bowker2010}. Inversion turns our eyes towards the ubiquity of infrastructure, how those infrastructures are not only symbolic but also material, that classifications were the product of historical indeterminancy, and a practical politics of what to make visible and what to keep hidden.

\section{A Research Agenda for the Genealogy of Machine Learning Data}
\label{sec:agenda}
Contesting data infrastructures through a genealogical method demands a new research agenda which addresses several dimensions of that infrastructure. While the agency and accountability of individual actors is not to be discounted, a genealogical investigation should also situate the actions of dataset creators and data subjects within historical contingencies and organizational and institutional contexts. We outline here an emerging research agenda, structured around four key questions.

First, {\it how do dataset developers in machine learning research describe and motivate the decisions that go into their creation?} By beginning with the datasets and their associated documentation (e.g. conference proceedings and communications and dataset documentation), we treat the dataset itself as a text. Reading the dataset as a text can help illuminate the motivations, spoken and unspoken conventions of dataset construction, curation, and annotation. In an analogous project, \citep{Geiger2020} analyzed the data collection and annotation practices of over a hundred social computing articles analyzing Twitter data and found a lack of consistent standardized practices of documentation. Following this line of research, we are currently analyzing a heterogeneous set of machine learning datasets from with computer vision using both structured and unstructured content analysis methods. In this interrogation, we attempt to reassemble which elements treat the data as a first-class research object and which elements designate it as a necessary by-product of doing cutting edge machine learning research. We also engage with texts via a grounded theory approach, by allowing themes and discourses to emerge inductively, rather than imposing a pre-established structure upon them.

This leads to our second research question: {\it what are the histories and contingent conditions of creation of benchmark datasets in machine learning?} Datasets, like all technical artifacts, have contingent and contextual social histories. Data which are gathered from individuals and stored in perpetuity in large-scale datasets have historical tendrils which are connected through those individuals and beyond them into scientists, technicians, and the artifacts which reify them. Datasets also bear marks of the matrix of power which shapes the relationship between scientist and patient, the same way HeLa cells were extracted from Henrietta Lacks, a Black woman cells whose cervical cancer cells were removed from her without knowledge of consent before her death in 1951 by white cell biologist George Ott Gey \citep{Skloot2011}. A genealogy of machine datasets ought to be retrospectively attentive to these histories and the ways in which the datasets themselves have been incorporated into the black box of regular machine learning practice.  Asking this question necessitates a deep dive into a handful of authoritative datasets by interpreting their histories and interviewing their creators and others who have labored upon them.

Third, {\it how do benchmark datasets become authoritative and how does this impact research practice?} The mass adoption of a dataset or a method, or other artifact or result does not stand alone. Just because there are dramatic improvements to a result does not automatically guarantee that it will be adopted more widely. Scientists who develop new tools and methods must enlist relevant literature, endure trials of skepticism by counter-laboratories, and mobilize allies by translating their interests into the interests of others \citep{latour1987science}. 
The centralization of research agendas around a small set of authoritative datasets is often accompanied by value-laden disciplinary commitments. For example, the emergence of the deep learning era, sparked in large part by ImageNet, has both necessitated and instigated increases in compute power, larger datasets, and specialized hardware -- components which are only possible to obtain within large tech companies and major research universities \citep{Dotan2020ValueladenDS}.

The convergence upon deep learning has analogues into many past large breakthroughs in technology and science; these analogues have been interrogated by historians and sociologists of science. \citet{kohler1994lords} asks how some types of organisms -- for instance, the {\it drosophila} mentioned above -- became {\it the} model organism for a particular field of study. Likewise, \citet{fujimura1988molecular} describes how molecular biology research was not driven by the force of the subdiscipline's applicability towards cancer research but was due to bandwagonning effects within the field. A similar type of effect may be at work within deep learning and the paradigmatic datasets associated with the move to deep learning. In this research agenda -- understanding that certain datasets are paradigmatic -- it's necessary to analyze the citation patterns, institutional and organizational networks, and research practice associated with several authoritative benchmark machine learning datasets.

Lastly, we ask {\it what are the current work practices, norms, and routines that structure data collection, curation, and annotation of data in machine learning?} The retrospective and historical methodologies that structure our previous three research questions provide important, but partial, perspectives on the current data practices within machine learning. The negotiations, norms, and assumptions that shape the creation of a dataset are often lost in the process of creating it, enmeshed in the practices and with no archival record.

Thus, our final research question aims to understand work practices {\it in situ}, by performing a multi-sited ethnography centered around the major computer science hubs that have contributed to the data infrastructure underlying current machine learning work, such as Silicon Valley (e.g. Stanford, Berkeley), Toronto (e.g. UofT, Vector), or Montreal (e.g. MILA). Treating major computer science labs as ethnographic sites will provide us with first-hand exposure to the work practices, negotiated transactions, and assumptions which undergird the creation of these datasets. Our work will build upon  growing ethnographic work focused on data science and machine learning teams \citep{passi2019problem, sachs2019algorithm, seaver2019captivating} and on a larger tradition of laboratory ethnography \citep{latour2013laboratory}.

\section{Conclusion}

Our goals in pursuing this research agenda are as follows. First, we want to develop a framework for data scientists and machine learning practitioners to reflexively analyze elements of their data pipeline which must be questioned and clarified before ever gathering a byte of data. Thinking about data within a dataset must be holistic, future-looking, and aligned with ethical principles and values. Datasets should be released not only with their technical specifications but additionally include a clear formulation of their stated objectives and the methodologies of collection, curation, and classification. In this sense, we echo the call of our colleagues who have done significant work around model and data transparency \citep{Gebru2018, Mitchell2019ModelCF}.


Second, we aim to push AI ethics conversation about data beyond issues associated with insufficient training data as the sole "solution" to racist, sexist, homophobic, and transphobic outcomes in sociotechnical systems. Gathering more training data from populations which are already extensively surveilled ignores how data-gathering operations reinscribe forms of domination and can serve as another form of "predatory inclusion". In this respect, the legitimate goal of data transparency that we referenced above should not be construed as a justification to place vulnerable populations in a visibility trap.  

Third, examining datasets moves the onus of our scientific inquiry away from people who are overwhelming the objects of data collection. Thus, we reverse the order of inquiry from data subjects to the creators of data collection, their taxonomic choices, decisions and their intended or unintended effects within a network of relations operative in a given dataset. In this respect, we choose to "studying up" by pointing our social scientific tools towards those with economic, social, and technological power to understand how their norms, values, and practices and power relations shape the data which undergirds everyday sociotechnical systems \citep{Nader1972UpTA, Forsythe99, Barabas2020}. 

Finally, this project points towards understanding the role of interrogating the invisible and undervalued labor plays in that goes into the construction development of datasets which amount - as we will see- to critical infrastructures for the development of machine learning. A growing literature with science and technology studies and anthropology of sociotechnical systems has focused on the importance of analyzing interrogating unspoken work practices of technical experts \citep{passi2019problem, seaver2019captivating}, subject matter experts in acts of data repair \citep{sachs2019algorithm}, and crowd laborers \citep{irani2013turkopticon, salehi2015we, gray2019ghost}. Interrogating those work practices and the politics of labor surrounding them forces us to articulate practices of accountability and contestability in the development of benchmark datasets.

\bibliography{main}
\bibliographystyle{icml2020}

\end{document}